\documentclass[12pt]{article}

\usepackage{latexsym}
\usepackage{amsmath}
\usepackage{amsfonts}
\usepackage{amssymb}
\usepackage{amsthm}
\usepackage{epsfig}
\usepackage{graphics}
\usepackage{natbib}
\usepackage{authblk}

\usepackage{setspace}
\usepackage{graphicx}
\usepackage{multirow}
\usepackage{float}
\usepackage{caption}
\usepackage{subcaption}
\usepackage[symbol]{footmisc}
\usepackage{xcolor}

\providecommand{\keywords}[1]
{
  \small	
  \textbf{\textit{Keywords---}} #1
}

\begin{document}
\title{Group Sequential Methods for the Win Ratio}
\author[1,*]{Tracy Bergemann} \author[2]{Tim Hanson}
\affil[1]{BridgeBio Pharma, Inc., Palo Alto CA 94304}
\affil[2]{Medtronic, Plc, Minneapolis MN 55432}

\affil[*]{Corresponding author: Tracy Bergemann, tracy.bergemann@bridgebio.com}

\maketitle

\newpage 

\textbf{Acknowledgments} None

\textbf{Artificial intelligence} No LLMs or generative AI were used in this manuscript.

\textbf{Declaration of conflicting interest} No conflicts with the research to declare.

\textbf{Funding statement}  None

\textbf{Ethical approval and informed consent statements} The IN.PACT SFA protocols were approved by the institutional review boards or ethics committees at each trial site. All patients provided written informed consent before enrollment. Trials were conducted in accordance with the Declaration of Helsinki, good clinical practice guidelines, and applicable laws as specified by all relevant governmental bodies.

\textbf{Data availability statement} Simulated data can be re-created using the simulation code provided in the Supplemental material and using the same seed. Data from the IN.PACT SFA clinical trial cannot be shared due to regulatory constraints that forbid secondary use of patient data from some participating countries.

\newpage 

\abstract{
\noindent The win ratio is increasingly used in randomized trials due to its intuitive clinical interpretation, ability to incorporate the relative importance of composite endpoints, and its capacity for combining different types of outcomes (e.g. time-to-event, binary, counts, etc.) to be combined. There are open questions, however, about how to implement adaptive design approaches when the primary endpoint is a win ratio, including in group sequential designs. A key requirement allowing for straightforward application of classical group sequential methods is the independence of incremental interim test statistics. This paper derives the covariance structure of incremental U-statistics that evaluate the win ratio under its asymptotic distribution. The derived covariance shows that the independent increments assumption holds for the asymptotic distribution of U-statistics that test the win ratio. Simulations confirm that traditional $\alpha$-spending preserves Type I error across interim looks.  A retrospective look at the IN.PACT SFA clinical trial data illustrates the potential for stopping early in a group sequential design using the win ratio. We have demonstrated that straightforward use of Lan-De\uppercase{M}ets $\alpha$-spending is possible for randomized trials involving the win ratio under certain common conditions. Thus, existing software capable of computing traditional group sequential boundaries can be employed.}


\section{Introduction}

Group sequential designs are commonly used in pivotal clinical trials to optimize the sample size requirements needed to demonstrate safety and efficacy.  The $\alpha$-spending framework of \cite{LD1983} controls for the overall type I error of the trial.  Several robust and widely-used statistical packages are available to compute decision boundaries, power, sample size requirements, etc., using the Lan-DeMets approach, e.g. the \texttt{gsDesign} package for R \citep{gsDesign}.  To apply group sequential methods with this framework, a multivariate integration of the joint density function of sequential test statistics is required.  An assumption is made that the increments used in the tests statistics are independent, simplifying the integration. Intuitively, the independent increment assumption holds for study outcomes that are collected at a fixed point in time.  For accruing data on longitudinal or time-to-event outcomes, however, the increments that contribute to the test statistics at each interim analysis are not necessarily independent.  For example, data collected on a time-to-event outcome in an event-driven study will continue to accrue information per patient over the increments such that they are correlated with past information.  
Previous research has shown that the asymptotic distribution for test statistics based on increments of longitudinal or time-to-event data meet the independence assumption.  The discussion in \cite{KT2020} thoroughly reviews this topic. The Lan-DeMets $\alpha$-spending is now widespread in clinical trials with these outcomes.  

More recently, test statistics of wins and losses on composite outcomes have gained popularity.  These tests include the Finkelstein-Schoenfeld (FS) test \citep{FS1999}, generalized pairwise comparisons \citep{B2010}, the win ratio \citep{PACW2012}, and the win odds adaptation of the win ratio by \cite{BVM2021}.  In each of these tests, several endpoints contribute to the composite and the components are typically ranked in a hierarchy of clinical severity. A treatment group is compared to a control group via an algorithm that determines who wins or loses in each pairwise comparison between the groups.  These statistics confer several advantages as a means to evaluate composite outcomes, including the ability to prioritize more severe events over less severe events, the capability to combine endpoints that are different types of metrics (e.g. continuous, time-to-event, and discrete), and provide a natural way to handle missing data for some or all of the components.

\cite{WRvsHR} compared the win ratio to traditional composite time-to-first-event survival endpoints across 16 studies.  They found that win ratio statistics were roughly comparable in magnitude and significance to treatment effects reported as hazard ratios obtained from the Cox proportional hazard model. \cite{WRvsHR} noted that, unlike traditional composite endpoints, the win ratio allows the prioritization of endpoints in order of importance and facilitate the inclusion of various metrics such as quality of life into the primary outcome.  \cite{Dong2020a} showed that in simplifying situations the win ratio reduces to the hazard ratio when proportional hazards hold for survival data, and reduces to the odds ratio for binary data.  \cite{Dong2020a} further introduced a modified win ratio termed the win odds that incorporates ties.

It is an open question of whether group sequential designs involving win-loss test statistics can be routinely employed.  Thus far, it is unclear whether the independent increment assumption can be satisfied in order to justify use of a Lan-DeMets alpha-spending function. Under an exact bivariate distribution of incremental test statistics, the increments are correlated. This is because evaluation of win-loss statistics is based on all pairwise comparisons between groups, and thus increments are not independent because new patients enrolled will be compared to pre-existing patients that have already contributed to comparisons. The open question is whether the independent increment assumption is met for the asymptotic distribution of incremental win-loss test statistics.

In \cite{FS1999}, the authors discuss the adaptation of their test statistic to sequential monitoring and state that the independent increment assumption would not be met. Instead, they recommend a modification that employs a block stratification of patients depending on the order in which they are enrolled into the trial. The authors expand on this further in \cite{FS2022}. There they also note that block stratification requires adequate spacing between looks in order to obtain independence, a requirement that is not often feasible. An alternative approach is provided to estimate the correlation between incremental test statistics: either via direct estimation or using a bootstrap. The bootstrap estimate of the covariance is used to generate boundaries for sequential monitoring, but this can only be performed at the time of the interim look, and can not be pre-specified. A simulation demonstrates that type I error is preserved in an example clinical trial. Comprehensive proof for type I error preservation is not provided. 

In this paper, we will demonstrate that, under certain conditions, boundaries do not need to be generated empirically and rather that the Lan-DeMets $\alpha$-spending function can be employed when sample sizes are sufficiently large such that the asymptotic distributions can safely be assumed.  To date, one publication has explored the use of group sequential designs for the win odds statistic and demonstrated with mathematical derivation and empirical simulation that indeed alpha-spending functions can be employed and appropriately control the overall type I error \citep{NMK2022}.  Our research expands on that work and aims to show that the independent increment assumption can be met for incremental test statistics under the asymptotic distribution of U-statistics as described in \cite{BL2016}.  These U-statistics can be used to test the win difference (also known as net benefit) or the win ratio.

Section 2 introduces a motivating example.  Section 3 defines the methodological framework and provides the mathematical derivation to show that the independent increment assumption is met for tests of the win difference and win ratio when only complete information is used for each subject.  Section 4 offers a set of simulations that demonstrate empirically that the overall type I error is controlled using typical $\alpha$-spending functions, both for the case when only complete patient information is used and for the case when complete and partial patient information is used at each interim analysis. Section 5 re-analyzes data from the IN.PACT SFA clinical trial, illustrating the potential for early stopping for success in a group sequential win ratio setting.  The paper concludes with a discussion in Section 6.

\section{Vascular disease example}
Vascular medicine trials commonly report patency as the primary endpoint of interest.  Patency is defined as a blood vessel that remains unobstructed and is usually informed by multiple sources of information, including whether the patient has retained their limb, the presence of wounds, whether an additional procedure was required to improve blood flow, and whether a diagnostic image shows blood flow through the treated vessel.  These data components are of different types and from different sources, such as right-censored events, binary or continuous variables from images collected at regular time intervals, or recurrent events.  A determination of wins and losses on a hierarchical composite can more easily combine information from these disparate types of data sources, and evaluate them according to a natural ranking that is clinically meaningful.

Consider a randomized controlled trial of patients with vascular disease where the lesion is treated with a novel therapy versus standard of care.  Patency is the primary endpoint of interest with hierarchical components consisting of (1) a major amputation, (2) a target lesion revascularization and (3) occlusion or restenosis of the vessel. Major amputation is a right-censored time-to-event variable, revascularization is a recurrent event, and occlusion or restenosis are measured longitudinally during regularly scheduled follow-up visits. The timing of events is also considered clinically relevant, such that if two patients both have a clinical event, the patient that experiences the event sooner is considered the loser in a win-loss determination. The majority of vascular disease trials evaluate patency as a primary endpoint through 6 months or 12 months. See, for example, \cite{Tepe2015} or \cite{Mustapha2019}.

The primary effectiveness endpoint can be assessed with a win ratio.  The win ratio parameter is $r$, where $r = \tau_1⁄\tau_2$  and $\tau_i$ is the probability that the subject in group $i$ is a winner for $i=1,2$.  Formally, the one-sided hypothesis to be tested is $H_0: r = 1$ versus $H_A: r \neq 1$.  The estimate of the win ratio is $\hat{r} = W/L$ where $W$ is the number of wins in the treatment group and $L$ is the number of losses in the treatment group for each pairwise comparison between the treatment group subjects and control group subjects.

\section{Methods}

Suppose the investigators plan to include a group sequential design to account for uncertainty in the estimate of the treatment effect.  Suppose that $K$ interim analyses are planned with a Lan-DeMets $\alpha$-spending function.  Interim analysis will occur at increments $k=1,\dots, K$ and stop for early success if $Z$-statistics surpass the boundaries. 

To avoid the dependence on the distribution of follow-up times as discussed in \cite{Oakes2016}, we suggest the trial is designed such that the statistical analysis of the primary endpoint is performed through time $T$, where the value of $T$ is the same for all participants. The framework is that patient data is collected for the primary endpoint at regularly scheduled intervals up to and including time $T$ and that information is used in analysis, but not information collected after time $T$. In other words, event-driven trials or trials where patient follow-up schedules are variable are not considered here. Patients with partial follow-up due to study dropout are appropriate to include as their status will not change with additional follow-up.  In those instances when a pairwise comparison includes a patient with incomplete follow-up, the follow-up of the other patient in the pair is truncated to match the minimum of the two follow-up times. Many trials used a fixed follow-up time $T$ at which to evaluate the primary endpoint, say patients evaluated based on a lab or image result at $T=6$ months, or evaluated for an event of interest through $T=12$ months. Examples include medical device trials such as PARTNER3 \citep{Mack2019}, REVERSE \citep{Linde2008} or ALLEVIATE-HF \citep{Butler2025} and pharmaceutical trials such as EMPULSE \citep{Biegus2023}, but there are an abundance of examples in the literature.

 The derivations in this section assume that only patients with completed follow-up are used in each interim evaluation and therefore information that each patient contributes to an interim analysis does not change from one interim analysis to the next. This constraint for interim analysis is also mentioned in a more recent paper by \cite{PSJ2024} when considering sample size re-estimation using promising zone methods. Patients are only included in an interim analysis if their data collection is final at follow-up time $T$, or have dropped out of the study before time $T$. This constraint is feasible to implement in trials where the primary endpoint is evaluated at a fixed time $T$ and $T \leq 12$ months. 
 
Suppose there are two samples randomly drawn $X_1, X_2, ..., X_m \sim F$ and $Y_1, Y_2, ..., Y_n \sim G$ for $i=1,...,m$ and $j=1,...,n$ and $N=n+m$ defines the total sample size.  Following \cite{BL2016}, the $U$-statistic is defined thusly
\[ U_{\nu} = \frac{1}{nm} \sum_{i=1}^m \sum_{j=1}^n \phi_{\nu}(X_i,Y_i), \hspace{0.1in} \nu=1,2,\] 
where
\begin{eqnarray*} \phi_1(X,Y) = I(X>Y) & \mbox{$X$ wins} \\
\phi_2(X,Y)= I(X<Y) & \mbox{$Y$ wins}
\end{eqnarray*}  Note that here we use $X>Y$, $X<Y$, and $X=Y$ to denote $X$ winning, losing, or tying $Y$, respectively, which is a hierarchical decision process for each pair.

Let $\tau_{\nu} = E\{\phi_{\nu}(X,Y)\}$.  Then
\[ \sqrt{N} \left[ \begin{array}{c} U_1 - \tau_1 \\ U_2 - \tau_2 \end{array} \right] \stackrel{\mathcal{D}}{\rightarrow} N_2 \left( \left[ \begin{array}{c} 0 \\ 0 \end{array} \right], \left[ \begin{array}{cc} \sigma_{11} & \sigma_{12} \\ \sigma_{12} & \sigma_{22} \end{array} \right] \right),\]
where
\[ \sigma_{uv} = \frac{N}{m} \xi_{10}^{uv} + \frac{N}{n} \xi_{01}^{uv}, \hspace{0.1in} u=1,2, \hspace{0.1in} v=1,2,\]
\[ \xi_{10}^{uv} = \mbox{cov}(\phi_u(X_1,Y_1),\phi_v(X_1,Y_1'),\]
and
\[ \xi_{01}^{uv} = \mbox{cov}(\phi_u(X_1,Y_1),\phi_v(X_1',Y_1).\]
Here and elsewhere $(X_1,X_1')$ and $(Y_1,Y_1')$ are pairs of outcomes from two different subjects.  The win ratio $\psi = \tau_1/\tau_2$, estimated by $\hat{\psi} = U_1/U_2$, has the asymptotic distribution
\[ \log( U_1/U_2) \stackrel{\mathcal{D}}{\rightarrow} N\left(\log(\psi), \frac{\sigma_{11}}{\tau_1}+\frac{\sigma_{22}}{\tau_2}-2 \frac{\sigma_{12}}{\tau_1 \tau_2} \right).\]

\subsection{Independent increments for Win difference}

Consider a group sequential design with test statistics $Z_k$ evaluated over time for $k=1,\dots, K$.  Consider two test statistics $Z_k$ and $Z_l$ computed from $m_k$ and $m_l$ subjects in the treatment group and $n_k$ and $n_l$ subjects in the control group at the two time points $k$ and $l$, with $k<l$.  In our calculations, it is assumed that patients contributing to test statistics will have complete information available, so that patient $i$ contributing information to $Z_k$ will also contribute the same information to $Z_l$.
To demonstrate that the independent increment assumption is met, it is sufficient to show that for two incremental test statistics $Z_k$ and $Z_l$ that under the asymptotic distribution $\mbox{cov}(Z_k, Z_l) = \sqrt{I_k(\theta) / I_l(\theta)}$. This is discussed in Equation 2.8 of \cite{KT2020}. 

First, in the case of U-statistics of the win difference, it can be shown that if $\mbox{cov}(U_{1k} - U_{2k} , U_{1l} - U_{2l}) = \mbox{var}(U_{1l} - U_{2l}) = 1/I_l(\theta)$ then 
$\mbox{cov}(Z_k, Z_l)  = \sqrt{I_k(\theta)}/\sqrt{I_l(\theta)}$. 

This can be shown in a straightforward manner as follows: 
\begin{eqnarray*}
\mbox{cov}(Z_k, Z_l) & = &
\mbox{cov}\left( \frac{U_{1k} – U_{2k} – (\tau_1 - \tau_2)}{\sqrt{\mbox{var}(U_{1k} – U_{2k})}}, \frac{U_{1l} – U_{2l} – (\tau_1 - \tau2)}{\sqrt{Var(U_{1l} – U_{2l})}} \right) \\
& = & \frac{\mbox{cov}(U_{1k} – U_{2k} , U_{1l} – U_{2l} }{\sqrt{Var(U_{1k} – U_{2k})}\sqrt{Var(U_{1l} – U_{2l})} } \\
& = & \sqrt{I_k(\theta)  I_l(\theta)} \mbox{cov}(U_{1k} – U_{2k} , U_{1l} – U_{2l} )
\end{eqnarray*}
Then, Proposition 1 shows that the condition is met.

\begin{flushleft}
\textbf{Proposition 1}:  For two incremental test statistics of the win difference, $U_{1k} - U_{2k}$ and $U_{1l} - U_{2l}$, \[\mbox{cov}(U_{1k}-U_{2k} , U_{1l}-U_{2l}) = \mbox{var}(U_{1l} -U_{2l}).\]
\end{flushleft}  
Proof is shown in the Appendix.

\subsection{Independent increments for Win ratio}

The results demonstrated for the win difference above are then extended to the win ratio.  For ease of derivation, and given that the asymptotic results shown in \cite{BL2016} also rely on the log transformation of the win ratio, the log transformation is applied here as well.

\begin{flushleft}
\textbf{Proposition 2}:  For two incremental test statistics of the log win ratio, $\log \left( \tfrac{u_{1k}}{u_{2k}} \right)$ and $\log \left( \frac{u_{1l}}{u_{2l}} \right)$,
\[\mbox{cov} \left( \log \left( \frac{u_{1k}}{u_{2k}} \right), \log \left( \frac{u_{1l}}{u_{2l}} \right) \right) \approx \mbox{var}\left[ \log \left( \frac{u_{1l}}{u_{2l}} \right) \right].\]
\end{flushleft}  
Proof is shown in the Appendix.

\section{Simulation Study}

\begin{figure}
    \centering
    \includegraphics[width=1\linewidth]{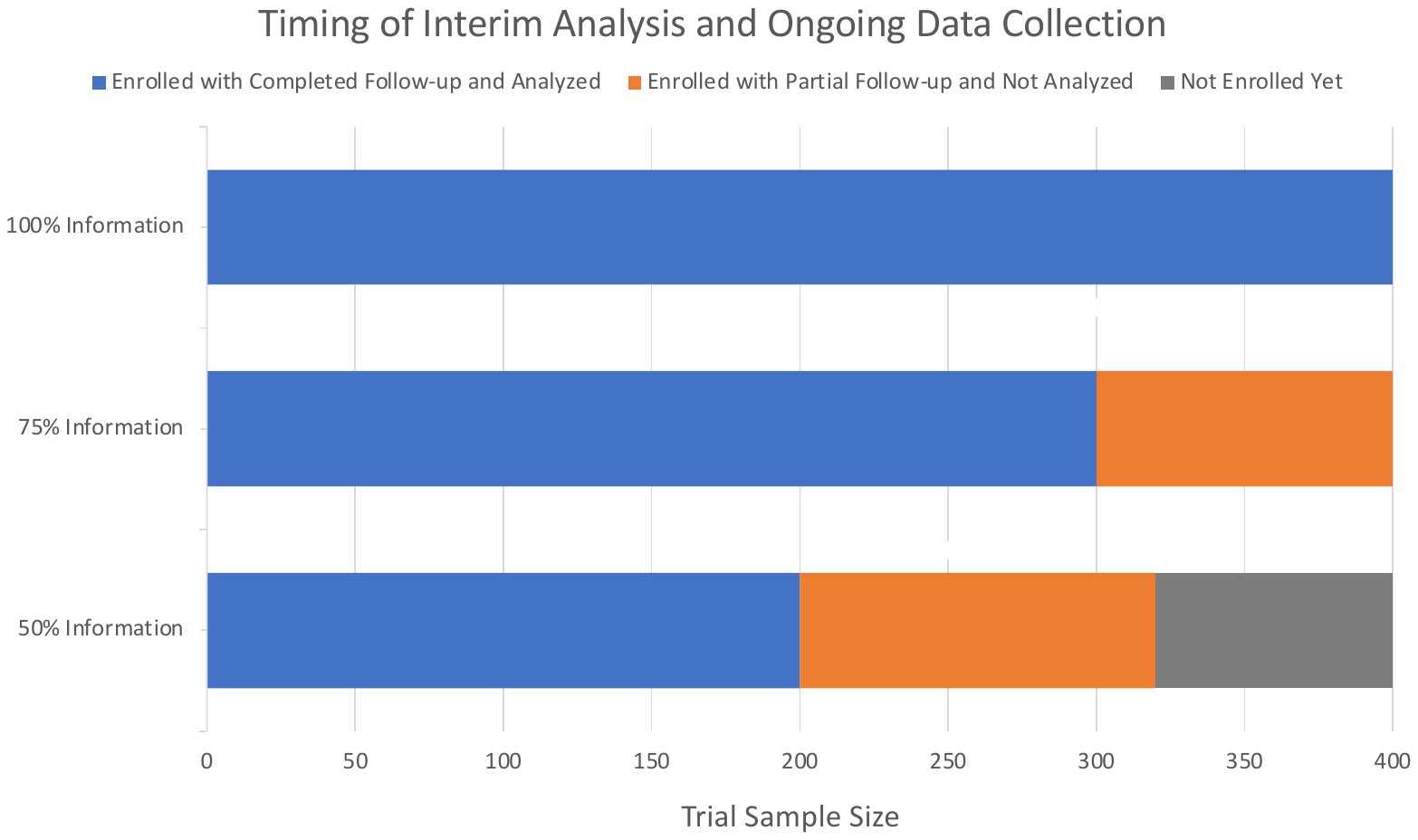}
    \caption{ }
    \label{fig:iatimes}
\end{figure}

Simulations were conducted to examine the conditions of the trial presented in the motivating example above (Scenario 1).  The conditions of the simulation are as follows.  Patients were randomized one-to-one to receive either treatment or control and are followed through $T=12$ months.  The rates of major amputation, target lesion revascularization, and occlusion through 12 months are assumed to be 3\%, 25\% and 55\% respectively.  It is further assumed that 40\% of target lesion revascularizations are precluded by a total occlusion that is detected by imaging at a follow-up visit.  To investigate the empirical estimate of overall type I error in the trial, rates are assigned to be the same in each randomization arm so that there is no treatment effect between the two arms.  Two trial sample sizes were chosen, $N=200$ and $N=400$. Figure \ref{fig:iatimes} illustrates which data is evaluated at each interim analysis in a generic simulated trial.

Code developed in base R 4.5.0 was used to evaluate the matrix of wins and losses for each pairwise comparison in order to estimate the win-ratio and derive the p-values from the test statistic.  The Hwang-Shih-DeCani $\alpha$-spending function with a gamma parameter of -3 is used to control the overall alpha level at 0.05 for a two-sided hypothesis test. The boundary condition at the interim analysis was based on the information fraction at the interim analysis and computed by dividing the number of subjects analyzed at the interim analysis by the total number of subjects at the final analysis. The boundaries were generated using the gsDesign library \citep{gsDesign} in R.

Further simulations were conducted to extend the motivating example to the case where patients could contribute partial follow-up information at interim analysis (Scenario 2).  In this extension, the information fraction at the interim analysis is then computed by dividing the number of person-years of follow-up analyzed at the interim analysis by the total person-years of follow-up at the final analysis.  The R code that was used to generate and evaluate simulated data is included as online supplementary material.

\subsection{Simulation results}

The results in Table 1 show that overall type I error is controlled in this empirical assessment of 10,000 simulated trials, even for fairly small sample sizes.  Under Scenario 1, when the total trial sample size is $N=200$, only 0.0517 of trials reject the null hypothesis at the first, second or final analysis.  Similarly when the sample size is $N=400$, only 0.0487 of trials reject the null hypothesis at any planned analysis, demonstrating effective control of type I error.  Similar results are upheld under Scenario 2 when partial information is also used.

\section{Example from the IN.PACT SFA Randomized Trial}

\begin{figure}
    \centering
    \includegraphics[width=1\linewidth]{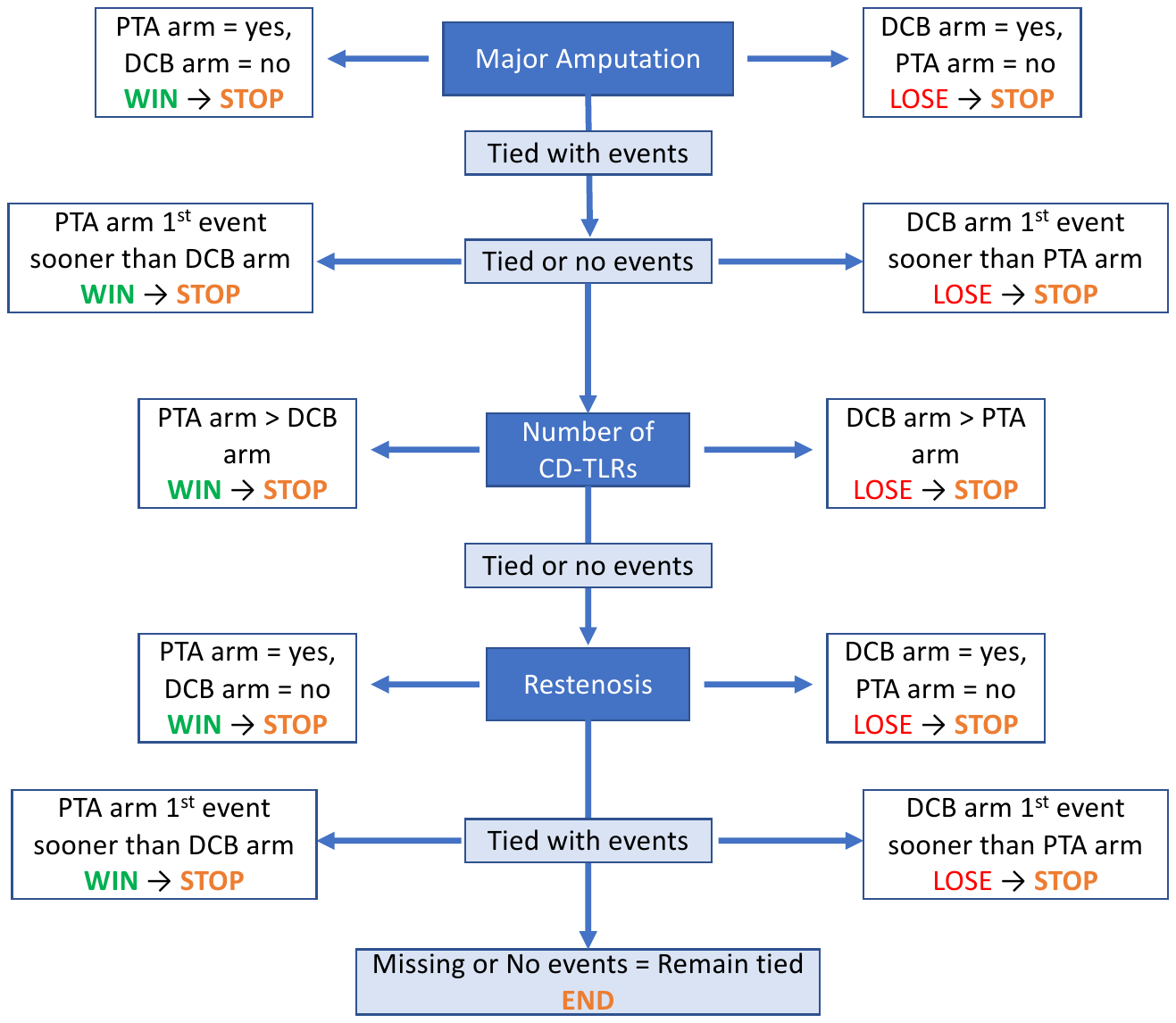}
    \caption{Determination of Wins and Losses for Patency Endpoint}
    \label{fig:flowchart}
\end{figure}

The IN.PACT SFA Randomized Trial was a multicenter, single-blinded, randomized trial with 331 patients randomly assigned in a 2:1 ratio to treatment with a drug-coated balloon (DCB) or percutaneous transluminal angioplasty (PTA) \citep{Tepe2015}. Patients presented with superficial femoral and popliteal peripheral artery disease. The primary endpoint for the trial was primary patency, defined as freedom from restenosis or clinically driven target lesion revascularization at 12 months. Figure \ref{fig:flowchart} shows how wins and losses are evaluated for patency. The trial met its primary endpoint with a difference between the primary patency in the DCB versus the PTA arm of 82.2\% versus 52.4\% (P $<$ 0.001). When the 12 month results from the trial are re-analyzed with the win ratio instead, a hierarchical endpoint is established with (1) major amputation, (2) number of clinically driven target lesion revascularizations and (3) restenosis. The win ratio for this endpoint is 3.82 with a 95\% confidence interval of (2.39, 6.10) and a p-value $<$ 0.001. This re-analysis with the win ratio instead shows results that are in line with the original publication.

Now suppose hypothetically, that a group sequential design had been employed in this trial. Suppose that interim analysis were scheduled when 50\% and 75\% of information was available for analysis, akin to the approach described in the Simulation section above. If an interim analysis had been performed when 50\% of information was available, that is, when 110 DCB and 56 PTA patients had completed 12 months of data, the win ratio endpoint at that time would have been 4.29 with a p-value $<$ 0.001.  Table 2 shows the win ratio and p-values at each interim look and the final. If a group sequential design had been employed in the IN.PACT SFA trial, the trial would have stopped for success at the first interim look.

\section{Conclusion}

This research has shown that the independent increment assumption is met for successive test statistics derived under the asymptotic distribution of the win difference in a group sequential design. The derivation was then extended to show that a similar result approximately holds for incremental tests of the win ratio.  Given that the derivation used for the win ratio tests relied on an approximation, these results were then supported with a simulation study to show that indeed type I error is well-preserved. Previous research has shown a similar result for win odds statistics \citep{NMK2022}.
 
 Our derivations hold for the case when only patients with completed follow-up are used in each evaluation and therefore information that each patient contributes to an interim analysis does not change from one interim analysis to the next.  It is of keen interest to see if the results we have derived here can be extended to situations where partial patient information is used in earlier interim analyses akin to the log-rank test.  We have conducted a simulation exercise to suggest that this may be the case, but further research is needed to establish that result.

\bibliography{references}
 
\section{Appendix}

\subsection{Proof of Proposition 1}

Consider the difference in wins and losses using the $U$-statistics defined in Bebu and Lachin (2016).  Consider two increments in an interim analysis, $k$ and $l$, out of $K$ total increments.  There are $m_k$ and $n_k$ subjects in increment $k$, and $m_l$ and $n_l$ patients in increment $l$. It is assumed that the information that a subject contributes to a test statistic is the same across all increments.  This means that $\phi_1(X_i,Y_j)$ and $\phi_2(X_i,Y_j)$ for any $i,j$ are the same from one increment to the next. Let \[\rho_{kl} = \mbox{cov}(U_{1k}-U_{2k},U_{1l}-U_{2l})\] be the covariance between the win/loss difference at two different interim looks.

Then
\begin{eqnarray*} \rho_{kl} & = & \mbox{cov}\left(\frac{1}{n_k m_k} \sum_{i=1}^{m_k} \sum_{j=1}^{n_k} \phi_1(X_i,Y_j)-\phi_2(X_i,Y_j),\frac{1}{n_l m_l} \sum_{i=1}^{m_l} \sum_{j=1}^{n_l} \phi_1(X_i,Y_j)-\phi_2(X_i,Y_j)  \right) \\ & = & 
\frac{1}{n_k m_k} \frac{1}{n_l m_l}  \left[  \sum_{i=1}^{m_k} \sum_{j=1}^{n_k} \mbox{cov} \left( \phi_1(X_i,Y_j), \sum_{i=1}^{m_l} \sum_{j=1}^{n_l} \phi_1(X_i,Y_j)-\phi_2(X_i,Y_j) \right) \right. \\ & & -\sum_{i=1}^{m_k} \sum_{j=1}^{n_k} \mbox{cov} \left. \left( \phi_2(X_i,Y_i), \sum_{i=1}^{m_l} \sum_{j=1}^{n_l} \phi_1(X_i,Y_j)-\phi_2(X_i,Y_j) \right) \right] \\ & = & 
\frac{1}{m_k n_k} \frac{1}{m_l n_l} \sum_{i=1}^{m_k} \sum_{j=1}^{n_k} \left[ \mbox{var}(\phi_1(X_i,Y_j)) \right. \\ & &
+ (n_l-1) \xi_{10}^{11} + (m_l-1) \xi_{01}^{11} - \mbox{cov}(\phi_1(X_i,Y_j),\phi_2(X_i,Y_j)) \\ & &
-(n_l-1) \xi_{10}^{12}-(m_l-1) \xi_{01}^{12} - \mbox{cov}(\phi_2(X_i,Y_j),\phi_1(X_i,Y_j)) \\ & & 
-(n_l-1) \xi_{10}^{21} -(m_l-1) \xi_{01}^{21} + \mbox{var}(\phi_2(X_i,Y_j)) \\ & & \left.
+(n_l-1) \xi_{10}^{22} + (m_l-1) \xi_{01}^{22} \right] \\  \mbox{(WLOG)} & = &
\frac{1}{m_l n_l} \left[ \mbox{var} (\phi_1(X_1,Y_1))+(n_l-1) \xi_{10}^{11} \right.  \\ & &
+(m_l-1) \xi_{01}^{11} - \mbox{cov}(\phi_1(X_1,Y_1),\phi_2(X_1,Y_1)) \\ & &
-(n_l-1) \xi_{10}^{12}-(m_l-1) \xi_{01}^{12} \\ & &
-\mbox{cov}(\phi_2(X_1,Y_1),\phi_1(X_1,Y_1))-(n_l-1) \xi_{10}^{21} \\ & &
-(m_l-1) \xi_{01}^{21} +\mbox{var}(\phi_2(X_1,Y_1)) \\ & &
\left. (n_l-1) \xi_{10}^{22} + (m_l-1) \xi_{01}^{22} \right] \\ & = &
\mbox{cov}(u_{1l},u_{1l})-\mbox{cov}(u_{1l},u_{2l}) -\mbox{cov}(u_{2l},u_{1l})+\mbox{cov}(u_{2l},u_{2l}) \\ & = &
\mbox{cov}(u_{1l}-u_{2l},u_{1l}-u_{2l}) \\ & = &
\mbox{var}(u_{1l}-u_{2l})
\end{eqnarray*}
Thus, the independent increment assumption is met under the asymptotic distribution of win difference test statistics according to the conditions described in \cite{KT2020}.  

\subsection{Proof of Proposition 2}

Our heuristic proof of independent increments for the win ratio test statistics uses the following approximation
\[ \mbox{cov}\left(\log(u_1),\log(u_2)\right) \approx \log \left[\frac{E(u_1 u_2)}{E(u_1) E(u_2)}+1 \right].\]
This approximation is derived from Taylor's expansion, the details of which are provided below.

Let \[ \rho_{kl}=\mbox{cov} \left( \log \left( \frac{u_{1k}}{u_{2k}} \right), \log \left( \frac{u_{1l}}{u_{2l}} \right) \right) \] be the covariance between the log win-ratio at two different interim looks. 

Then
\begin{eqnarray*}
\rho_{kl} & = & \mbox{cov} \left[ \log(u_{1k})-\log(u_{2k}),\log(u_{1l})-\log(u_{2l}) \right] \\ 
& = & \mbox{cov} [ \log(u_{1k}),\log(u_{1l})]-\mbox{cov}[\log(u_{1k}),\log(u_{2l}) ] \\
& & - \mbox{cov}[\log(u_{2k}),\log(u_{1l})] +\mbox{cov}[\log(u_{2k}),\log(u_{2l}) ] \\
& \approx & \log \left[ \frac{\mbox{cov}(u_{1k},u_{1l})}{\tau_1 \tau_2} + 1 \right]-\log \left[ \frac{\mbox{cov}(u_{1k},u_{2l})}{\tau_1 \tau_2} + 1 \right] \\
& & -\log \left[ \frac{\mbox{cov}(u_{2k},u_{1l})}{\tau_1 \tau_2} + 1 \right] + \log \left[ \frac{\mbox{cov}(u_{2k},u_{2l})}{\tau_1 \tau_2} + 1 \right] \\
& = & \log \left[ \frac{\mbox{cov}(u_{1l},u_{1l})}{\tau_1 \tau_2} + 1 \right]-\log \left[ \frac{\mbox{cov}(u_{1l},u_{2l})}{\tau_1 \tau_2} + 1 \right] \\
& & -\log \left[ \frac{\mbox{cov}(u_{2l},u_{1l})}{\tau_1 \tau_2} + 1 \right] + \log \left[ \frac{\mbox{cov}(u_{2l},u_{2l})}{\tau_1 \tau_2} + 1 \right] \\
& \approx & \mbox{cov} [ \log(u_{1l}),\log(u_{2l})]-\mbox{cov}[\log(u_{1l}),\log(u_{2l})] \\
& & - \mbox{cov}[\log(u_{2l}),\log(u_{1l})] +\mbox{cov}[\log(u_{2l}),\log(u_{2l})]  \\
& = & \mbox{cov} [ \log(u_{1l})-\log(u_{2l}), \log(u_{1l})-\log(u_{2l}) ] \\ 
& = & \mbox{var}\left[ \log \left( \frac{u_{1l}}{u_{2l}} \right) \right]
\end{eqnarray*}

The approximation used in the win ratio uses the Taylor's moment expansion 
\[E(f(X)) \approx f(\mu_x) + \tfrac{1}{2} f''(\mu_x)\mbox{var}(X).\]

Using $f(x) = \exp(x)$ and the Taylors approximation $\exp\left(\tfrac{1}{2} \mbox{var}(X) \right) \approx 1 + \tfrac{1}{2} \mbox{var}(X)$
 yields
\[E[\exp(X)] = \exp(\mu_x)+\tfrac{1}{2} \exp(\mu_x)\mbox{var}(X) \approx \exp\left(\mu_x + \tfrac{1}{2} \mbox{var}(X) \right).\] 
Thus,
\begin{eqnarray*}
1 + \frac{\mbox{cov}(u_1,u_2)}{E(u_1)E(u_2)} & = & \frac{\mbox{cov}(u_1,u_2)+E(u_1) E(u_2)}{E(u_1)E(u_2)} \\ 
& = & \frac{E(u_1 u_2) -E(u_1)E(u_2) +E(u_1)E(u_2)}{E(u_1)E(u_2)} \\
& = & \frac{E(u_1 u_2)}{E(u_1)E(u_2)} \\
& \approx & \frac{\exp\left(\mu_a + \mu_b + \tfrac{1}{2} \mbox{var}(a+b)\right)}{\exp\left(\mu_a + \tfrac{1}{2} \mbox{var}(a)\right) \exp\left(\mu_b + \tfrac{1}{2} \mbox{var}(b)\right)} \\
& \approx & \exp\left(\mbox{cov}(a,b) \right)
\end{eqnarray*}
where $a = \log(u_1)$ and $b=\log(u_2)$.

\begin{table}
\label{tab1}
\caption{Simulation Type I Error and Win Ratio Estimates for $N=200$ and $N=400$.}
\begin{tabular}{lllll}
& First interim & Second Interim & Final analysis & Reject $H_0$ \\
& 50\% information & 75\% information & 100\% information & at any analysis \\ \hline \hline
\multicolumn{3}{l}{Scenario 1: Complete information only} & & \\ \hline \hline
$N=200$ & & & & \\ \hline
Proportion & 0.0082 & 0.0135 & 0.0300 & 0.0517 \\
rejecting $H_0$ & & & & \\ \hline
WR mean & 1.0518 & 1.0353 & 1.0274 & \\
WR median & 1.0024 & 1.0069 & 1.0039 & \\ \hline \hline
$N=400$ & & & & \\ \hline
Proportion & 0.0084 & 0.0135 & 0.0268 & 0.0487 \\
rejecting $H_0$ & & & & \\ \hline
WR mean & 1.0244 & 1.0158 & 1.0120 & \\
WR median & 1.0061 & 1.0032 & 1.0045 & \\ \hline \hline
\multicolumn{3}{l}{Scenario 2: Complete and partial information} & & \\ \hline \hline
$N=200$ & & & & \\ \hline
Proportion & 0.0091 & 0.0098 & 0.0235 & 0.0444 \\
rejecting $H_0$ & & & & \\ \hline
WR mean & 1.0277 & 1.0276 & 1.0274 & \\
WR median & 1.0029 & 1.0035 & 1.0039 & \\ \hline \hline
$N=400$ & & & & \\ \hline
Proportion & 0.0094 & 0.0090 & 0.0230 & 0.0414 \\
rejecting $H_0$ & & & & \\ \hline
WR mean & 1.0130 & 1.0125 & 1.0120 & \\
WR median & 1.0044 & 1.0039 & 1.0045 & \\ \hline \hline\end{tabular}
\end{table}

\begin{table}
\label{tab2}
\caption{Interim Analysis in the IN.PACT SFA trial}
\begin{tabular}{llll}
& First interim & Second Interim & Final analysis \\
& 50\% information & 75\% information & 100\% information \\ \hline \hline
\hline
Win ratio & 4.288 & 3.650 & 3.823 \\
p-value & $<$ 0.0001 &  $<$ 0.0001 & $<$ 0.0001 \\ \hline
Nominal boundary & 0.0091 & 0.0177 & 0.0413 \\ 
Signficance & Yes & Yes & Yes \\
\hline \hline
\end{tabular}
\end{table}

\end{document}